\documentclass[reprint,superscriptaddress,showpacs,nofootinbib,onecolumn]{revtex4-1}
\usepackage{setspace}
\usepackage[dvips]{graphicx}
\usepackage{amsmath,amssymb,amsthm,mathrsfs,amsfonts,dsfont}
\usepackage{subfigure, epsfig}
\usepackage{braket}
\usepackage{bm}
\usepackage{enumerate}
\usepackage{graphicx}   
\usepackage{color}
\usepackage[10pt]{moresize}
\usepackage{amscd}
\usepackage{epsfig}
\usepackage{subfigure}
\usepackage{graphicx}
\usepackage{multirow}
\usepackage{verbatim}
\usepackage[outdir=./]{epstopdf}

\begin{document}

\title {Testing Local Realism into the Past without Detection and Locality Loopholes }

\author{Ming-Han Li}
\affiliation{Shanghai Branch, National Laboratory for Physical Sciences at Microscale and Department of Modern Physics, University of Science and Technology of China, Shanghai 201315, P.~R.~China}
\affiliation{Shanghai Branch, CAS Center for Excellence and Synergetic Innovation Center in Quantum Information and Quantum Physics, University of Science and Technology of China, Shanghai 201315, P.~R.~China}

\author{Cheng Wu}
\affiliation{Shanghai Branch, National Laboratory for Physical Sciences at Microscale and Department of Modern Physics, University of Science and Technology of China, Shanghai 201315, P.~R.~China}
\affiliation{Shanghai Branch, CAS Center for Excellence and Synergetic Innovation Center in Quantum Information and Quantum Physics, University of Science and Technology of China, Shanghai 201315, P.~R.~China}

\author{Yanbao Zhang}
\affiliation{ NTT Basic Research Laboratories and NTT Research Center for Theoretical Quantum Physics, NTT Corporation, 3-1 Morinosato-Wakamiya, Atsugi, Kanagawa 243-0198, Japan}

\author{Wen-Zhao Liu}
\affiliation{Shanghai Branch, National Laboratory for Physical Sciences at Microscale and Department of Modern Physics, University of Science and Technology of China, Shanghai 201315, P.~R.~China}
\affiliation{Shanghai Branch, CAS Center for Excellence and Synergetic Innovation Center in Quantum Information and Quantum Physics, University of Science and Technology of China, Shanghai 201315, P.~R.~China}

\author{Bing Bai}
\affiliation{Shanghai Branch, National Laboratory for Physical Sciences at Microscale and Department of Modern Physics, University of Science and Technology of China, Shanghai 201315, P.~R.~China}
\affiliation{Shanghai Branch, CAS Center for Excellence and Synergetic Innovation Center in Quantum Information and Quantum Physics, University of Science and Technology of China, Shanghai 201315, P.~R.~China}

\author{Yang Liu}
\affiliation{Shanghai Branch, National Laboratory for Physical Sciences at Microscale and Department of Modern Physics, University of Science and Technology of China, Shanghai 201315, P.~R.~China}
\affiliation{Shanghai Branch, CAS Center for Excellence and Synergetic Innovation Center in Quantum Information and Quantum Physics, University of Science and Technology of China, Shanghai 201315, P.~R.~China}

\author{Weijun Zhang}
\affiliation{State Key Laboratory of Functional Materials for Informatics, Shanghai Institute of Microsystem and Information Technology, Chinese Academy of Sciences, Shanghai 200050, P.~R.~China}

\author{Qi Zhao}
\affiliation{Center for Quantum Information, Institute for Interdisciplinary Information Sciences, Tsinghua University, Beijing 100084, P.~R.~China}

\author{Hao Li}
\affiliation{State Key Laboratory of Functional Materials for Informatics, Shanghai Institute of Microsystem and Information Technology, Chinese Academy of Sciences, Shanghai 200050, P.~R.~China}

\author{Zhen Wang}
\author{Lixing You}
\affiliation{State Key Laboratory of Functional Materials for Informatics, Shanghai Institute of Microsystem and Information Technology, Chinese Academy of Sciences, Shanghai 200050, P.~R.~China}

\author{W. J. Munro}
\affiliation{ NTT Basic Research Laboratories and NTT Research Center for Theoretical Quantum Physics, NTT Corporation, 3-1 Morinosato-Wakamiya, Atsugi, Kanagawa 243-0198, Japan}

\author{Juan Yin}
\author{Jun Zhang}
\author{Cheng-Zhi Peng}
\affiliation{Shanghai Branch, National Laboratory for Physical Sciences at Microscale and Department of Modern Physics, University of Science and Technology of China, Shanghai 201315, P.~R.~China}
\affiliation{Shanghai Branch, CAS Center for Excellence and Synergetic Innovation Center in Quantum Information and Quantum Physics, University of Science and Technology of China, Shanghai 201315, P.~R.~China}

\author{Xiongfeng Ma}
\affiliation{Center for Quantum Information, Institute for Interdisciplinary Information Sciences, Tsinghua University, Beijing 100084, P.~R.~China}

\author{Qiang Zhang}
\author{Jingyun Fan}
\author{Jian-Wei Pan}
\affiliation{Shanghai Branch, National Laboratory for Physical Sciences at Microscale and Department of Modern Physics, University of Science and Technology of China, Shanghai 201315, P.~R.~China}
\affiliation{Shanghai Branch, CAS Center for Excellence and Synergetic Innovation Center in Quantum Information and Quantum Physics, University of Science and Technology of China, Shanghai 201315, P.~R.~China}

\begin{abstract}
Inspired by the recent remarkable progress in the experimental test of local realism, we report here such a test that achieves an efficiency greater than $(78\%)^2$ for entangled photon pairs separated by 183 m. Further utilizing the randomness in cosmic photons from pairs of stars on the opposite sides of the sky for the measurement setting choices, we not only close the locality and detection loopholes simultaneously, but also test the null hypothesis against local hidden variable mechanisms for events that took place 11 years ago (13 orders of magnitude longer than previous experiments). After considering the bias in measurement setting choices, we obtain an upper bound on the $p$ value of $7.87\times10^{-4}$, which clearly indicates the rejection with high confidence of potential local hidden variable models. One may further push the time constraint on local hidden variable mechanisms deep into the cosmic history by taking advantage of the randomness in photon emissions from quasars with large aperture telescopes.
\end{abstract}

\maketitle

It has long been known that many of the predictions of quantum mechanics are counterintuitive and are strictly prohibited by local realism, our usual model of the world. This led to the famous question \textquotedblleft Can a Quantum-Mechanical Description of Physical Reality be Considered Complete?\textquotedblright~by Einstein, Podolsky, and Rosen in 1935 \cite{PhysRev.47.777}. Local realism requires that a system possesses an exact property prior to its measurement and the cause-effect action is limited by the speed of light. Quantum mechanics on the other hand presents a different description of our world, by allowing the presence of quantum superposition and nonlocal correlations between distant entangled particles. These non-local quantum mechanical correlations provide predications incompatable with local realism. John Bell introduced his celebrated inequality for a definitive hypothesis test of local realism to end the dispute~\cite{bell1964instein,bell2004speakable}.

Bell considered that the two parties Alice and Bob make a joint measurement on their remotely separated entangled particles. For measurement setting choices $x$, $y$ $\in\{0,1\}$, they receive the measurement outcomes $a$, $b$ $\in\{0,1\}$, respectively. According to local hidden variable models, the outcomes $a$ and $b$ are completely (pre)determined for the inputs $x$, $y$ and a hidden variable $\lambda$ carrying the exact state information such that $a=a(x,\lambda)$ and $b=b(y,\lambda)$. Local hidden variable models set a bound on the joint measurement probability distribution $p(a,b|x,y,\lambda)$, while quantum mechanical predictions surpass this bound~\cite{bell1964instein,PhysRevLett.23.880,PhysRevD.10.526,bell2004speakable}. The experimental violation of the Bell inequality was observed shortly after its derivation, and is now routinely performed in quantum physics laboratories (see~\cite{RevModPhys.86.419,larsson2014loopholes,PhysRevA.93.032115} for a recent review). However, the imperfections in experiments open loopholes for local hidden variable theories to reproduce the observed violation of Bell inequality which would otherwise be a strong evidence against local realism.

The detection of entangled particles in a Bell test experiment can be corrupted by loss and noise. The consequence of this is that the ensemble of the detected  states may not be an honest representative of what the source actually emits. It was shown that the local hidden variable models can explain the observed violation of Bell inequality if the detection efficiency of single entangled particles is $\leq2/3$~\cite{PhysRevA.47.R747}, which is known as the detection (fair sampling) loophole~\cite{PhysRevD.2.1418}. The Bell test experiment requires one to separate the measurement setting choice and measurement outcome on one side space-like from the measurement setting choice on the other. Failure to do so opens the locality loophole~\cite{jarrett1984physical}, which allows the two parties to communicate about their measurement settings before outputting outcomes. Further, the Bell test experiment also requires the measurement setting choices to be \textquotedblleft truly free or random~\textquotedblright, and that \textquotedblleft they are not influenced by the hidden variables\textquotedblright~\cite{bell2004speakable}. The lightcones of the measurement events on both sides and the entanglement creation at the source in a Bell test experiment cross each other in the past direction. A hidden cause in the common past can manipulate the experimental outcomes in the Bell test experiment, opening the freedom-of-choice loophole~\cite{bell2004speakable}.

The experimental test of Bell inequality was pioneered by Freedman and Clauser \cite{PhysRevLett.28.938}, and Aspect, Grangier, Dalibard and Roger \cite{PhysRevLett.47.460,PhysRevLett.49.91,PhysRevLett.49.1804}. The  locality  and detection loopholes were individually closed initially by Aspect et al.~\cite{PhysRevLett.49.1804}, Weihs et al.~\cite{PhysRevLett.81.5039}, and Rowe et al.~\cite{rowe2001experimental} followed by a number of others ~\cite{ansmann2009violation,hofmann2012heralded,giustina2013bell,PhysRevLett.111.130406}. Scheidl et al. made the first attempt on the freedom-of-choice loophole~\cite{scheidl2010violation}. Several groups have recently succeeded in closing both locality and detection loopholes simultaneously in Bell test experiments~\cite{hensen2015loophole,PhysRevLett.115.250402,PhysRevLett.115.250401,PhysRevLett.119.010402}. Spacetime analysis shows that the common past in these experiments~\cite{PhysRevLett.115.250402,PhysRevLett.115.250401} began by less than $<10^{-5}$ s before the experiment. We denote this time as $t_{cm}$, with $t_{cm} =-10^{-5}$~s with respect to the starting time of the Bell experiment. The outcomes in these experiments are possibly subject to the influence of local hidden variable models taking place before $t_{cm}$. In this Letter, we report on achieving $t_{cm} = -11$ years by using the randomness in cosmic photons for measurement setting choices in a Bell test experiment. With both detection and locality loopholes closed, and considering the distribution bias in measurement settings observed in the experiment, the prediction-based ratio (PBR) analysis method~\cite{Knill:2017,zhang:2013,zhang:2011} produces a $p$ value upper bound for the null hypothesis test to be $\le 7.873 \times 10^{-4}$, indicating a rejection of local hidden variable models taking place after $t_{cm} = -11$~yrs with high confidence.

Shown from our experimental layout depicted in Fig. 1, we create entangled photon pairs at 1560 nm by a spontaneous parametric downconverion process (SPDC) periodically at a reptition rate of 2 MHz and distribute the two photons of each pair via single mode optical fiber in opposite directions to Alice and Bob, which are at a distance of 93 m and 90 m from the source, respectively. At each measurement station, the entangled photons exit the fiber and pass through a Pockels cell for the polarization state measurement. They are then coupled into the single mode optical fiber to be detected by the superconducting nanowire single-photon detectors (SNSPD)~\cite{zhang2017nbn}. The heradling efficiency of single photons, from creation to detection, is obtained as the ratio of the two-photon coincident counting events with respect to the single-photon counting events, which is $(78.8\pm1.9)\%$ for Alice and $(78.7\pm1.5)\%$ for Bob in the experiment~\cite{pereira2013demonstrating,PhysRevLett.120.010503}, is sufficient to close the detection loophole. We measure a two-photon quantum interference visibility of $99.4\%$ in the Horizontal/Vertical base and $98.4\%$ in the diagonal(anti-diagonal) base. Further we obtain a fidelity of $98.66\%$ in the quantum state tomography measurement, with which we numerically find the nonmaximally polarization-entangled state, $\cos(22.05^\circ)\ket{HV}+\sin(22.05^\circ)\ket{VH}$. With the measurement settings: $-83.5^\circ$ (for $x=0$) and $-119.4^\circ$ (for $x=1$) for Alice, and $6.5^\circ$ (for $y=0$) and $-29.4^\circ$ (for $y=1$) for Bob, we obtain an optimum violation of the Bell inequality~\cite{PhysRevA.47.R747}(see Supplemental Material for the detailed description of quantum state characterization).

\begin{figure*}[tbh]
\centering
\resizebox{16cm}{!}{\includegraphics{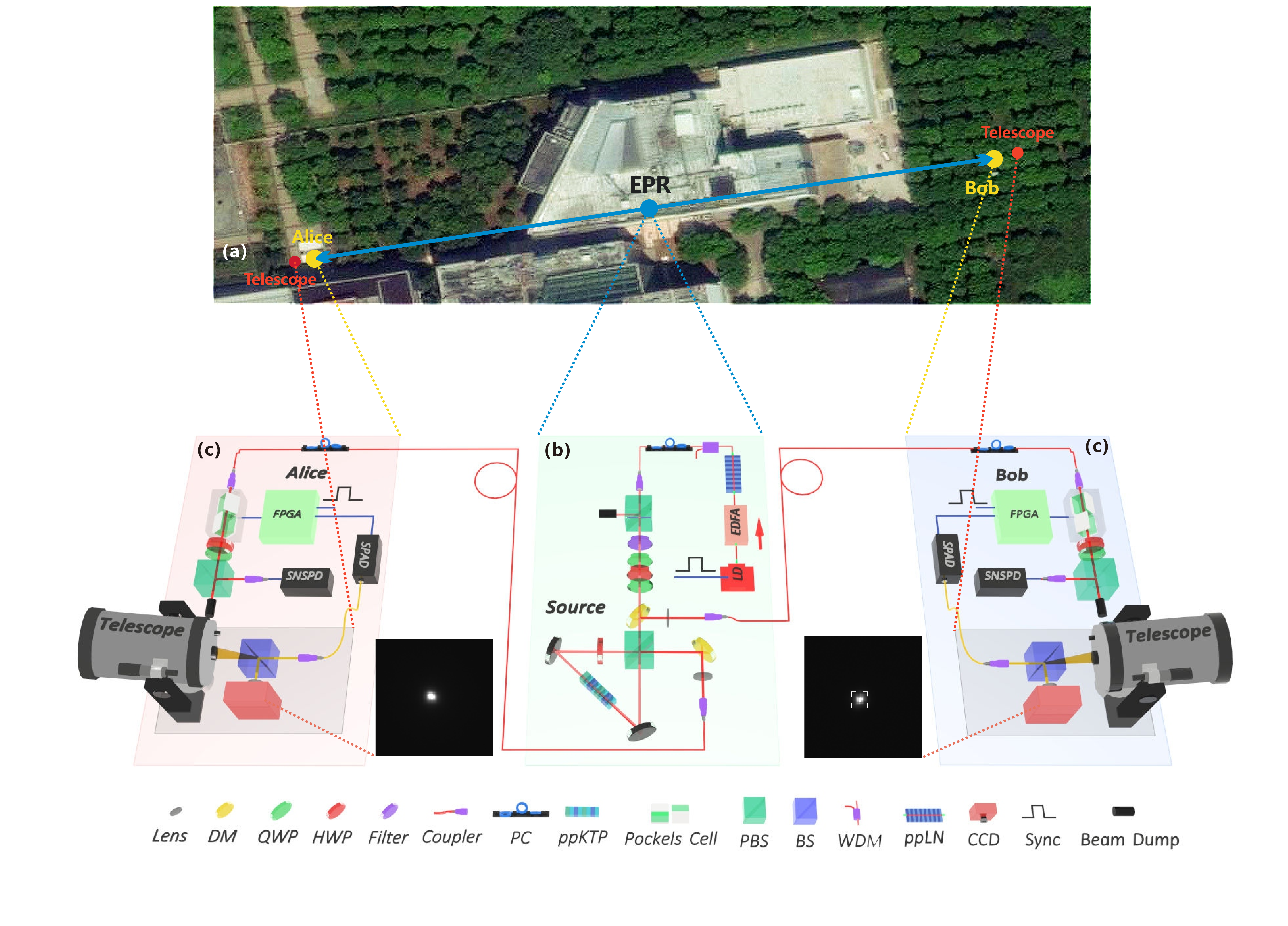}}
\caption{Experimental schematics. a) A bird's-eye view  of the experimental layout. Alice's and Bob's measurement stations are on the opposite sides of, and respectively 93$\pm$1 m and 90$\pm$1 m from the entangle photon pair source (labeled by EPR in the figure). Both Alice and Bob have a telescope $3$ m from the measurement station to collect cosmic photons.
b) Creation of pairs of entangled photons: light pulses of 10 ns, 2 MHz from a 1560 nm seed laser diode (LD) are amplified by an erbium-doped fiber amplifier (EDFA), and frequency-doubled in an in-line periodically poled lithium niobate (PPLN) crystal. With the residual 1560 nm light removed by a wavelength-division multiplexer (WDM) and spectral filters, the 780 nm light pulses are focused into a periodically poled potassium titanyl phosphate (PPKTP) crystal in a Sagnac loop to generate polarization entangled photon pairs. A set of quarter-wave plates (QWPs) and half-wave plate (HWP) are then used to control the relative amplitude and phase in the created polarization-entangled two-photon state. The residual 780 nm pump light is removed by dichroic mirrors (DMs). The two photons of an entangled pair at 1560 nm travel through optical fiber in opposite directions to two measurement stations, where they are subject to polarization state measurements.
c) Single photon polarization state measurement: the single photons exit the fiber, go through the polarization state measurement in free space, and are collected into a single mode optical fiber to be detected by superconducting nanowire single-photon detectors (SNSPDs). The apparatus to perform single-photon polarization measurement consists of a Pockels cell, QWP, HWP, and polarizing beam splitter (PBS). The cosmic photons collected by the telescope are split by a beam splitter (BS). The transmitted photons are coupled into the optical fiber and detected by a SPAD. The reflected photons form an image on a CCD camera, which is used for star tracking and stabilizing the coupling of cosmic photons into optical fiber. The SPAD outputs are fed to a field programmable gate array to generate random bits for measurement setting choice to trigger the Pockels cell to switch between two polarization orientations. A time-to-digital converter (TDC) is used to time-tag the events of cosmic random number generation and single-photon detection (see Supplemental Material for detailed experimental setup, which includes Refs.~\cite{zhang2017nbn}). (Insets) Star images (HIP 43813 and HIP 86032, respectively, for Alice and Bob) on the CCD camera for star tracking.}
\label{fig1}
\end{figure*}

Because of the presence of a common past, one can only test against local hidden variable models taking place after $t_{cm}$ in a Bell test experiment. Bell and a few others considered to use the randomness generated long before the experiment in the Universe to make $t_{cm}$ significantly large in the past direction~\cite{bell1976introductory,scheidl2010violation,maudlin2011quantum,vaidman2001tests}. Employing the randomness in certain properties of cosmic photons such as the arrival time, color, and polarization for the measurement setting choice in a Bell experiment has attracted significant recent attention~\cite{PhysRevLett.112.110405,PhysRevLett.118.060401,PhysRevLett.118.140402,leung2017astronomical}. Here we present a Bell test experiment employing the randomness in the creation time of cosmic photons. Therefore, the arrival times of a pair of cosmic photons, which are respectively emitted by a pair of cosmic sources located on the opposite sides of our sky, are random and so can be used for Alice's and Bob's measurement setting choices. To do so, at each measurement station, we use a telescope (Celestron, CPC 1100 HD) to receive photons from the selected cosmic radiation source, which has a diameter of 280 mm and a focal length of $f=2.94$ m. We use a beam splitter to reflect a portion of the collected cosmic photons to form an image of the cosmic source on a CCD camera (Andor Zyla, $2048\times2048$ pixels with a pexiel size of 6.5 $\mu m$) and couple the transmitted photons into a multimode fiber with $NA=0.22$ and a core diameter of 105 $\mu m$  which is connected to a silicon single photon avalanche diode (SPAD). We estimate the total detection efficiency of a single cosmic photon within the spectral band of silicon SPAD to be $<1\%$~\cite{PhysRevLett.118.140402}. Both the sensitive area of the CCD camera and the fiber end facet are at the focal plane of the telescope. The intensity profile of the image of the cosmic source is used in a tracking mechanism to stabilize the coupling of photons from the cosmic source into the fiber during the experiment~\cite{ren2017ground}. We pass the cosmic photon detection signals from the SPAD to a field programmable gate array (FPGA), which converts the random arrival time of cosmic photons into random bits for our Alice's and Bob's measurement setting choices.

\begin{figure}[tbh]
\centering
\resizebox{8cm}{!}{\includegraphics{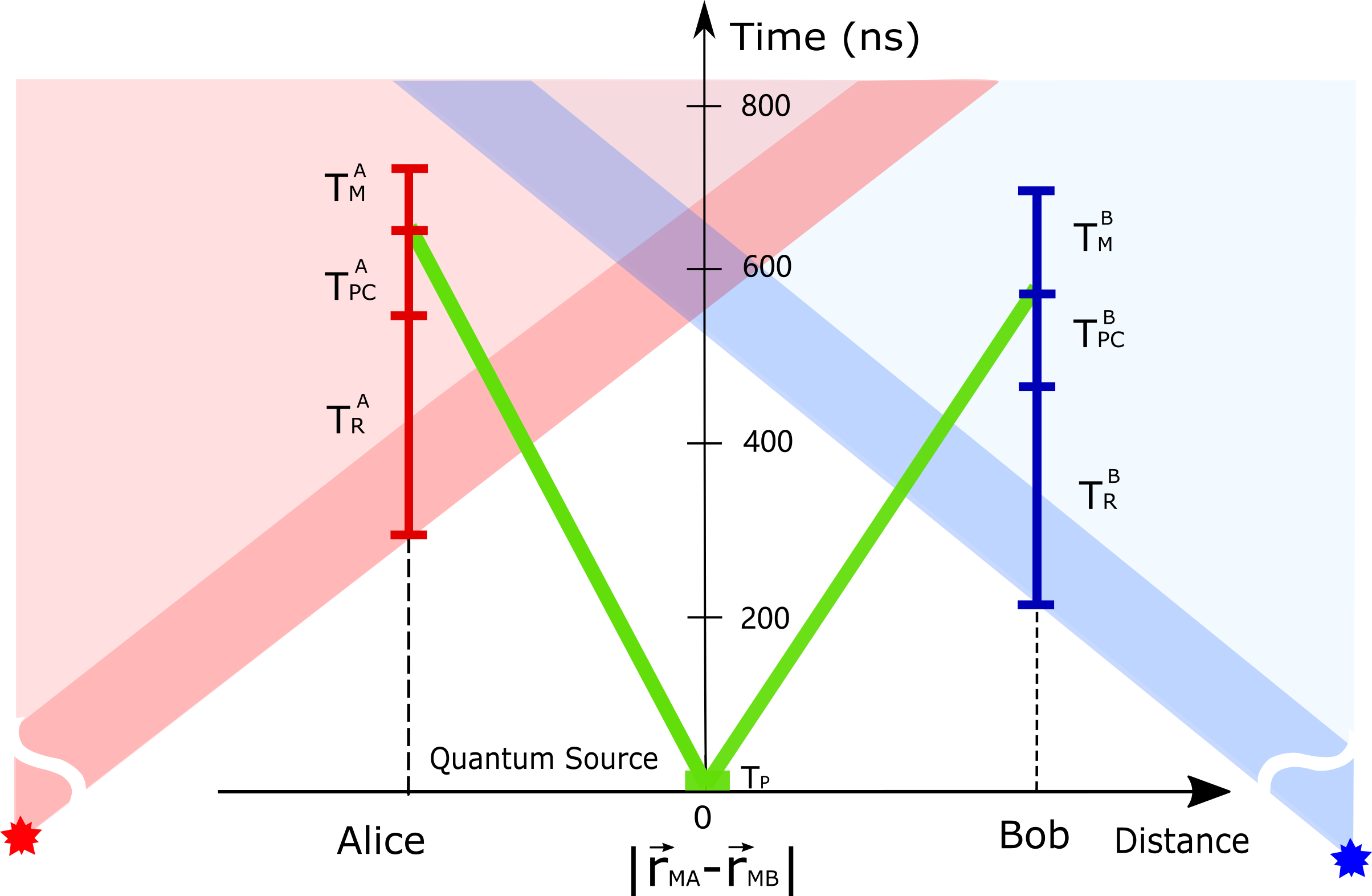}}
\caption{Spacetime diagram of the Bell test experiment. The green dot represents the event of creating an entangled photon pair in the source while the two thick green lines stand for delivering the photons to Alice and Bob via optical fiber, both having an uncertainty of $10$ ns, which is the temporal duration of the pump laser pulse. For Alice's (Bob's) side, the red (blue) line segments labeled by $T_R^A$ ($T_R^B$), $T_{PC}^A$ ($T_{PC}^B$), and $T_M^A$ ($T_M^B$) represent the time elapse starting from cosmic photons arriving at the telescope to the Pockels cell receiving a random bit, then to an entangled single photon leaving the Pockels cell, then to the photon detection circuit outputting a signal. The red (blue) strip stands for the time window to accept the cosmic photons for random bit generation, satisfying the spacelike separation condition through the entire duration of the experiment.}
\label{fig2}
\end{figure}

The spacetime diagram of our experiment is presented in Fig. 2 beginning with the event of creating a state in the source (at a repetition rate of 2 MHz) as the origin. To ensure that the measurement setting choice of the Alice (Bob) measurement is space-like separated from the measurement process of Bob (Alice) measurement, we require that when a cosmic photon from the selected cosmic source arrives at the telescope of Alice (Bob), its wave front has not arrived before Bob (Alice) finishes his (her) state measurement, which is quantified by two parameters, $\Gamma^{A}$ and $\Gamma^{B}$, respectively. Having $\Gamma^A >0$ and $\Gamma^B>0$ means we satisfy the space-like separation condition (See Supplemental Material for details about the derivation of $\Gamma^{A}$ and $\Gamma^{B}$, which includes Refs.~\cite{stone2011engineering, Meeus:1991:AA:532892, HIP}). In our experiment with the available choices of stars, by setting the time window to accept the cosmic photons to be 133.2 ns, we have $\Gamma^A>58$~ns and $\Gamma^B>60$~ns (as shown in Table I). It is important to note here that we only consider the optical refraction effect due to the atmosphere of Earth and assume that the interstellar space is vacuum in the current spacetime analysis. The interstellar medium has extremely low density~\cite{RevModPhys.73.1031, RevModPhys.54.1,Cordes1991,ISM}. It will be interesting to consider the possible delay of light propagation due to the interstellar medium in the future when the precise relevant information is available. Here, we assume that the propagation of cosmic photons and their arrival time are not affected by any means other than the known mechanisms in astronomy studies such as refraction through slowly varying interstellar medium, and assume the effect is identical for all photons.

\begin{table*}[htb]
    \centering
    \caption{Key parameters in the Bell test experiment. In each experimental run, we select a pair of stars with one for Alice and one for Bob, find the smallest ${t_{cm}}$ and $\Gamma_{A(B)}$ considering Earth's rotation, list the start time (UTC time) and run time~\cite{HIP}, the ratio between the number of random number \textquotedblleft0\textquotedblright~ and the number of random number \textquotedblleft1\textquotedblright~(0/1),~ signal-to-noise ratio (SNR), and the calculated bias. $\delta\Gamma_{A(B)} = 4$ ns.}
    \begin{tabular}{ccccccccccc}
   \hline
    Run	& &HIP ID&Start (UTC) time& Run time(s)& $t_{cm} \pm \delta_{t_{cm}}$ (yr)&$\Gamma_{A(B)}$ & \textquotedblleft0\textquotedblright/\textquotedblleft1\textquotedblright & SNR & Bias\\
    \hline
    1&	Alice&21421&2018/3/23 13h34m&416&	-36.71 $\pm$ 0.22&144&1.0078&491.6&0.00295\\
	&Bob&	69673A&&&&69&1.0012&94.7&0.00552\\
    \hline
    2&	Alice&	27989&2018/3/23 14h14m &2026&	-75.03 $\pm$ 3.73&	155&1.0053&584.6&0.00217\\
	&Bob&	76267&&&&			60&0.9989&111.6&0.00471\\
    \hline
    3&	Alice&	37279&2018/3/23 15h13m &2638&	-11.46 $\pm$ 0.05&	105&1.0059&147.4&0.00483\\
	&Bob&	80816&&&&				58&0.9985&78.5&0.00666\\
    \hline
    4&	Alice&	43813&2018/3/23 16h26m &2330&	-48.58 $\pm$ 0.77&	99&1.0009&125.2&0.00419\\
	&Bob&	86032&&&&				71&0.9987&132.6&0.00407\\
  \hline
\end{tabular}
\label{tab3}
\end{table*}

We use a master clock to produce synchronization signals at a repetition rate of 2 MHz, which is used to trigger the source to produce states. It further serves as an external clock to the FPGA. We convert the random arrival time of a cosmic photon received in the time window into a 1-bit random number in the following simple way: if a photon-detection signal from the SPAD is within 133.2 ns before a clock signal (at 2 MHz) in the FPGA, the FPGA outputs a random bit 0 if the photon-detection signal appears in the first 66.6 ns, while if the photon-detection signal appears in the second 66.6 ns window, the FPGA outputs a random bit 1. The FPGA does not output random bits in the following 5 $\mu$s, instead it applies an artificial 5 $\mu$s deadtime (see Supplemental Material for detailed description of synchronization and random number generation with cosmic photons). This cosmic random number generator outputs random bits at a rate not exceeding 200,000 $s^{-1}$, which is the maximum rate the Pockels cells can be switched at to realize the measurement settings in the current experiment. We now define an experimental trial as the case when both cosmic random number generators on the two sides simultaneously produce a random bit.

The local conditions, such as the weather (humidity, atmospheric turbulence, temperature variation), tall buildings, light pollution, besides the sky glow in Shanghai, permit us to select only among a few stars of low magnitude with our 280 mm telescopes for the experiment. Because of Earth's rotation and the time spent on finding the stars and optimizing the coupling of stellar photons into the fiber, we did the experiment with a pair of stars selected for Alice and Bob for about $10$-$40$ mins and then choose another pair of stars to continue the experiment. We conservatively set $t_{cm}=-11.46$~yrs (see Table I). The signal-to-noise ratio of the produced random numbers ranges from 78 to 584. The noise is taken by pointing the telescope slightly away from the cosmic source (at a dark patch of the sky). We notice that the ratio of the frequency of bit 1 with respect to that of bit 0 deviates slightly from the ideal value of 1, which may be due to the system imperfections including single photon detector deadtime and the time window broadening due to processing the detection of multi photons in the FPGA~\cite{JMO2009}, which will be further optimized in future work.

Noting that we have achieved high single-photon detection efficiency and ensured spacelike separation between relevant events, we now illustrate that we indeed close loopholes in our Bell test experiment, particularly the detection loophole, which has two related issues. One is related to the loss of entangled photons from creation to detection due to the system imperfection, and the other is related to the inefficiency in detecting cosmic photons. From a pedagogical perspective, we present the discussion with a general nonlocal game, where the two stars can be regarded as two referees. In each trial of the game, Alice and Bob, as two players who are not allowed to communicate during the trial, each receives a 1-bit random number as the input $x$ and $y$, and outputs a 1-bit outcome $a$ and $b$, respectively. The score for each trial is calculated according to the inputs and outputs. We stress that in each trial of the game, both referees give random bits, and the detection loophole problem arises when one or both players do not always have outputs (say, due to channel loss). In this case, Alice and Bob can prepare some (input and output) bits ahead and, if their input bits match the referees, they would output the corresponding output bits, and they would not provide outputs if the input bits are not matched. Such detection loophole was well studied in the past, which can be closed with high single-photon detection efficiency. On the other hand, it is okey that one or both referees sometimes do not want to play and therefore do not provide random bits in the game. We stress that it is not counted as a game trial when this happens; and it is counted as a game trial if and only if both referees provide random numbers at both sides. Hence, such cases do not introduce the detection loophole. We remark that  similar treatments have already been employed, e.g., in the spot-checking device-independent protocol \cite{miller2017universal,arnon2016simple}, in which only a small fraction of trials are randomly selected as test trials for the loophole-free Bell test, but the security of an information task based on other trial results is guaranteed.

We quantify the small bias in the generated random bits distribution by the total-variance distance from the uniform distribution. Because it is impossible to fully characterize this bias, we make the assumption that the random bits distribution bias at each measurement station in each trial is bounded to and independent of each other. We remark that under this situation we allow the measurement dependence, i.e., the dependence of the distribution of input random numbers at each measurement station on the local hidden variables as studied in Refs.~\cite{Hall2011, Barrett2011, Puetz:qc2014b}. We also allow the possibility that the distribution bias changes from trial to trial, and our data analysis method can take advantage of the knowledge of the bias change over time. The evidence against the null hypothesis of local realism, under the above assumption, is quantified in a reference computed using a test statistics. The $p$ value is the maximum probability according to local realism that the statistics takes a value as extreme as the observed one. Hence, a small $p$ value implies a strong evidence against local realism. We apply the PBR analysis method in designing the test statistics and computing an upper bound of the $p$ value. The PBR analysis was originally developed for the test of local realism, assuming that the input measurement setting distribution is fixed and known~\cite{zhang:2011, zhang:2013}, and later extended to the case with a relaxed assumption that the setting distribution bias is bounded~\cite{Knill:2017}. Hence, the PBR analysis method can be applied to our current situation. The PBR analysis provides valid upper bounds of $p$ values, even if the local realistic models depend on previous trial results and the experimental distribution of trial results varies over time. The bias in random bit distribution varies for different stars under the study, as shown in Table I. The PBR analysis incorporating the time-varying bias returns a $p$ value upper bound of $p=7.873\times10^{-4}$.  If we make a stronger but unjustified assumption that the measurement setting distribution is perfectly uniform, the PBR analysis returns a smaller $p$ value upper bound $p'=3.106\times 10^{-10}$ (see Supplemental Material for the detailed description of PBR analysis method, which includes Refs.~\cite{Knill:2017, zhang:2013, zhang:2011, Fine:1982, Peres:1999, PRBox, Barrett2005, Hoeffding}). Both indicate a rejection of the assumed local hidden variable models with high statistical confidence. Compared with the recent loophole-free Bell tests reported in~\cite{hensen2015loophole,PhysRevLett.115.250402,PhysRevLett.115.250401,PhysRevLett.119.010402}, our $p$ value upper bounds are larger than the $p$ value of $3.74\times 10^{-31}$ in~\cite{PhysRevLett.115.250401} or $2.57\times 10^{-9}$ in~\cite{PhysRevLett.119.010402}, and comparable to the $p$ value from $5.9\times 10^{-3}$ to $9.2\times 10^{-6}$ in ~\cite{PhysRevLett.115.250402}, but smaller than the $p$ value of 0.019 in~\cite{hensen2015loophole}.

In conclusion, we perform a null hypothesis test which rejects local hidden variable models taking place as early as 11 years before the experiment with high confidence. Looking into the future, our experiment may serve as a benchmark to progressively rule out local hidden variable models deep into the cosmic history by utilizing the randomness in quasars of high redshift or even cosmic microwave backgroud in future experiments. Further, we may find interesting applications in device-independent quantum information processing~\cite{pironio2010random,PhysRevLett.111.130406,bierhorst2017experimentally,PhysRevLett.120.010503,colbeck2012free,PhysRevLett.95.010503,pironio2009device,PhysRevLett.113.140501,gallego2013full}. Scaling up the spacetime extension in the local realism test is being actively pursued~\cite{Yin1140,PhysRevLett.120.140405}. The same system may also help to examine the hypothesis for human free choice~\cite{bell2004speakable,PhysRevLett.105.250404,RevModPhys.86.419,PhysRevA.93.032115,0264-9381-29-22-224011,Barrett2011,Hardy2017,hardy2017proposal} and gravitational effect~\cite{Penrose1996,DIOSI1987377} and to address collapse locality loophole~\cite{PhysRevA.72.012107,Greenberger2009,10.3389/fpsyg.2013.00845,PhysRevLett.103.113601}.

{\it Acknowledgment.---}  Y. Zhang would like to thank E. Knill for the stimulating discussions about the PBR analysis method. This work has been supported by the National Key R\&D Program of China (2017YFA0303900, 2017YFA0304000), the National Fundamental Research Program, National Natural Science Foundation of China, and Chinese Academy of Science.

M-H. L., C. W. and Y. Z. contributed equally to this work.

{\it Note added.---} Recently, we become aware of a related work~\cite{PhysRevLett.121.080403}.

\bibliography{Bibliography}

\clearpage

\onecolumngrid
\appendix

\setcounter{equation}{0}
\setcounter{table}{0}
\setcounter{figure}{0}

\centerline{\large\bf Supplemental Material:}
\vspace{0.15cm}
\centerline{\large\bf Test of Local Realism into the Past without Detection and Locality Loopholes}
\vspace{0.5cm}

\section{System characterization}

\subsection{Determination of single photon efficiency in the experimental system}
We define the single photon heralding efficiency as $\eta_A=C/N_B$ and $\eta_B=C/N_A$ for Alice and Bob, in which the coincidence events C and the single events for Alice $N_A$ and Bob ($N_B$) are measured in the experiment. The heralding efficiency is given by
\begin{equation} \label{Eq:heraldingEff}
\eta = \eta^{sc} \times \eta^{so} \times \eta^{fiber} \times \eta^{m} \times \eta^{det},
\end{equation}
where $\eta^{sc}$ is the efficiency to couple entangled photons into single mode optical fiber,  $\eta^{so}$ the efficiency for photons passing through the optical elements in the source, $\eta^{fiber}$ the transmittance of fiber linking the source to the measurement station, $\eta^{m}$ the efficiency for light passing through the measurement station, and $\eta^{det}$ the single photon detector efficiency. $\eta^{so}$, $\eta^{fiber}$, $\eta^{m}$, $\eta^{det}$ can be measured with classical light beams and NIST-traceable power meters. The coupling efficiency $\eta^{sc}$ is calculated with
\begin{equation} \label{Eq:heraldingEff}
\eta^{sc} = \frac{\eta}{\eta^{so} \times \eta^{fiber} \times \eta^{m} \times \eta^{det}},
\end{equation}

\begin{table}[htb]
\centering
  \caption{Characterization of optical efficiencies in the experiment. }
\begin{tabular}{c|c|ccccc}
\hline
 & heralding efficiency ($\eta$) & $\eta^{sc}$ & $\eta^{so}$ & $\eta^{fiber}$ & $\eta^{m}$ & $\eta^{det}$ \\
\hline
Alice  & 78.8\% & 93.9\% & \multirow{2}{*}{95.9\%} & \multirow{2}{*}{99\%} & 94.8\% & 93.2\%\\
Bob    & 78.7\%& 94.4\% &                       &                       & 95.2\% & 92.2\% \\
\hline
\end{tabular}
\label{tab:OptEffAB}
\end{table}

The transmittance of optical elements used in our experiment are listed in Table~\ref{tab:OptEff}, with which we obtain the efficiency $\eta^{so}$:
\begin{equation} \label{Eq:heraldingEff}
\eta^{so} = \eta^{AS} \times \eta^{S} \times (\eta^{DM})^4 \times \eta^{780/1560 HWP} \times \eta^{780/1560 PBS} \times \eta^{PPKTP} = 95.9\%,
\end{equation}
where we use four dichroic mirrors.

The transmittance of the 130 meter fibre between the source and the detection is $99\%$. The transmittance of the measurement station including the Pockels cell is $94.8\%$ for Alice and $95.2\%$ for Bob. The efficiency of the superconducting nanowire single-photon detector (SNSPD) \cite{zhang2017nbn} is measured to be $93.2\%$ for Alice and $92.2\%$ for Bob. The single photon heralding efficiency of the system is determined to be $\eta_A=(78.8\pm1.9)\%$ for Alice and $\eta_B=(78.5\pm1.5)\%$ for Bob with photon-counting statistic in the experiment.

\begin{table}[htb]
\centering
  \caption{The efficiencies of optical elements}
\begin{tabular}{c|c|c}
\hline
Optical element & Efficiency\\
\hline
$\eta^{AS}$ & Aspherical lens & $99.27\%\pm0.03\%$ \\
$\eta^{S}$ & Spherical lens & $99.6\%\pm1.0\%$ \\
$\eta^{780/1560 HWP}$ & Half wave plate (780nm/1560nm) & $99.93\%\pm0.02\%$ \\
$\eta^{1560 HWP}$ & Half wave plate (1560nm) & $99.92\%\pm0.04\%$ \\
$\eta^{1560 QWP}$ & Quarter wave plate (1560nm) & $99.99\%\pm0.08\%$ \\
$\eta^{780/1560 PBS}$ & Polarizing beam splitter (780nm/1560nm) & $99.6\%\pm0.1\%$ \\
$\eta^{1560 PBS}$ & Polarizing beam splitter (1560nm) & $99.6\%\pm0.2\%$ \\
$\eta^{DM}$ & dichroic mirror & $99.46\%\pm0.03\%$ \\
$\eta^{PPKTP}$ & PPKTP & $99.6\%\pm0.2\%$ \\
$\eta^{P}$ & Pockels cell & $98.7\%\pm0.5\%$ \\
\hline
\end{tabular}
\label{tab:OptEff}
\end{table}

\subsection{Quantum state characterization}
To maximally violate the Bell inequality in experiment, we create non-maximally entangled two-photon state $\cos(22.053^\circ)\ket{HV}+\sin(22.053^\circ)\ket{VH}$ (with $r=0.41$ for $(\ket{HV}+r\ket{VH})/\sqrt{1+r^2}$) and set the bases for single photon polarization state measurement to be  $A_1=-83.5^\circ$, $A_2=-119.38^\circ$ for Alice, $B_1=6.5^\circ$, $B_2=-29.38^\circ$ for Bob. We measure diagonal/anti-diagonal visibility in the bases set ($45^\circ, -22.053^\circ$), ($112.053^\circ, 45^\circ$) for minimum coincidence, and in the bases set ($45^\circ, 67.947^\circ$), ($22.053^\circ, 45^\circ$) for maximum coincidence, where the angles represent measurement basis  $cos(\theta)\ket{H}+sin(\theta)\ket{V}$ for Alice and Bob.

By setting the mean photon number to $\mu=0.0035$ to suppress the multi-photon effect, we measure the visibility to be $99.4\%$ and $98.4\%$ in horizontal/vertical basis and diagonal/anti-diagonal basis.

\begin{figure}[htb]
\centering
    \subfigure[]{
      \includegraphics[width=8cm]{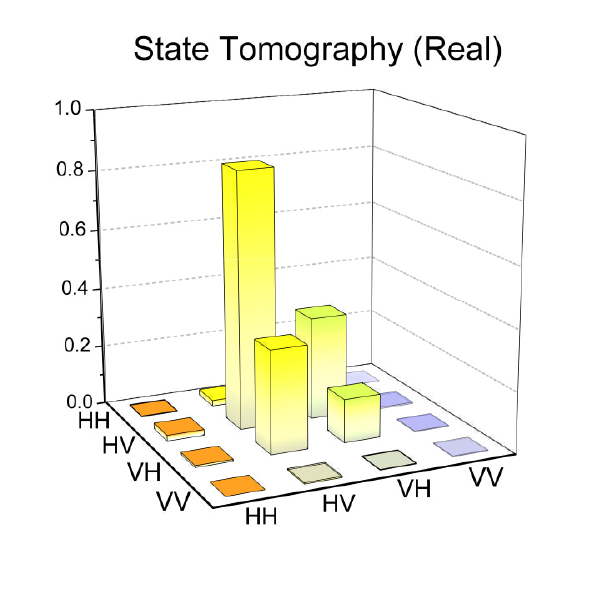}
    }
    \subfigure[]{
      \includegraphics[width=8cm]{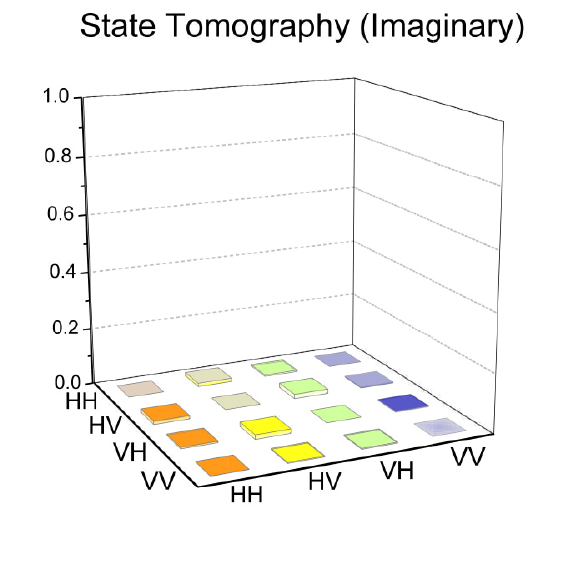}
    }
\caption{(color online)
Tomography of the produced state. The real and imaginary part are shown in (a) and (b).}
\label{Fig.Tomo}
\end{figure}

We perform state tomography on the non-maximally entangled state, the result is shown in Fig.~\ref{Fig.Tomo}, the state fidelity is $98.66\%$. We attribute the imperfection to multi-photon components, imperfect optical elements, and imperfect spatial/spectral mode matching.

\section{Spacetime analysis}
Because the randomness of cosmic photons are from the remote cosmic radiation sources, the Earth rotation must be considered in computing the spacetime relations of ground-based experimental sites and cosmic radiation sources.
Taking the event of generating an entangled photon pair in the source as the origin in a spacetime diagram, the positions of measurement station and the cosmic source are given as $\vec{r}_{MA}$ and $\vec{r}_{SA}$ for Alice's side, and $\vec{r}_{MB}$ and $\vec{r}_{SB}$ for Bob's side, respectively. We consider the scenario that when a cosmic photon from the selected cosmic radiation source arrives the telescope of Alice (Bob), its wavefront has not arrived before Bob (Alice) finishes the measurement on his (her) part of the entangled quantum state as shown in Fig.~\ref{fig1}, which can be described with the inequalities below

\begin{eqnarray}\label{eqa}
  \begin{split}
  \frac{|\vec{r}_{SA} - \vec{r}_{MB}|}{c} + t_{SA} > \frac{n\cdot L_{B}}{c} + T_{M}^B + T_{P}, \\
  \frac{|\vec{r}_{SB} - \vec{r}_{MA}|}{c} + t_{SB} > \frac{n\cdot L_{A}}{c} + T_{M}^A + T_{P},
  \end{split}
\end{eqnarray}
\begin{figure*}[tbh]
\centering
\includegraphics [width=11.39cm,height=8.2cm]{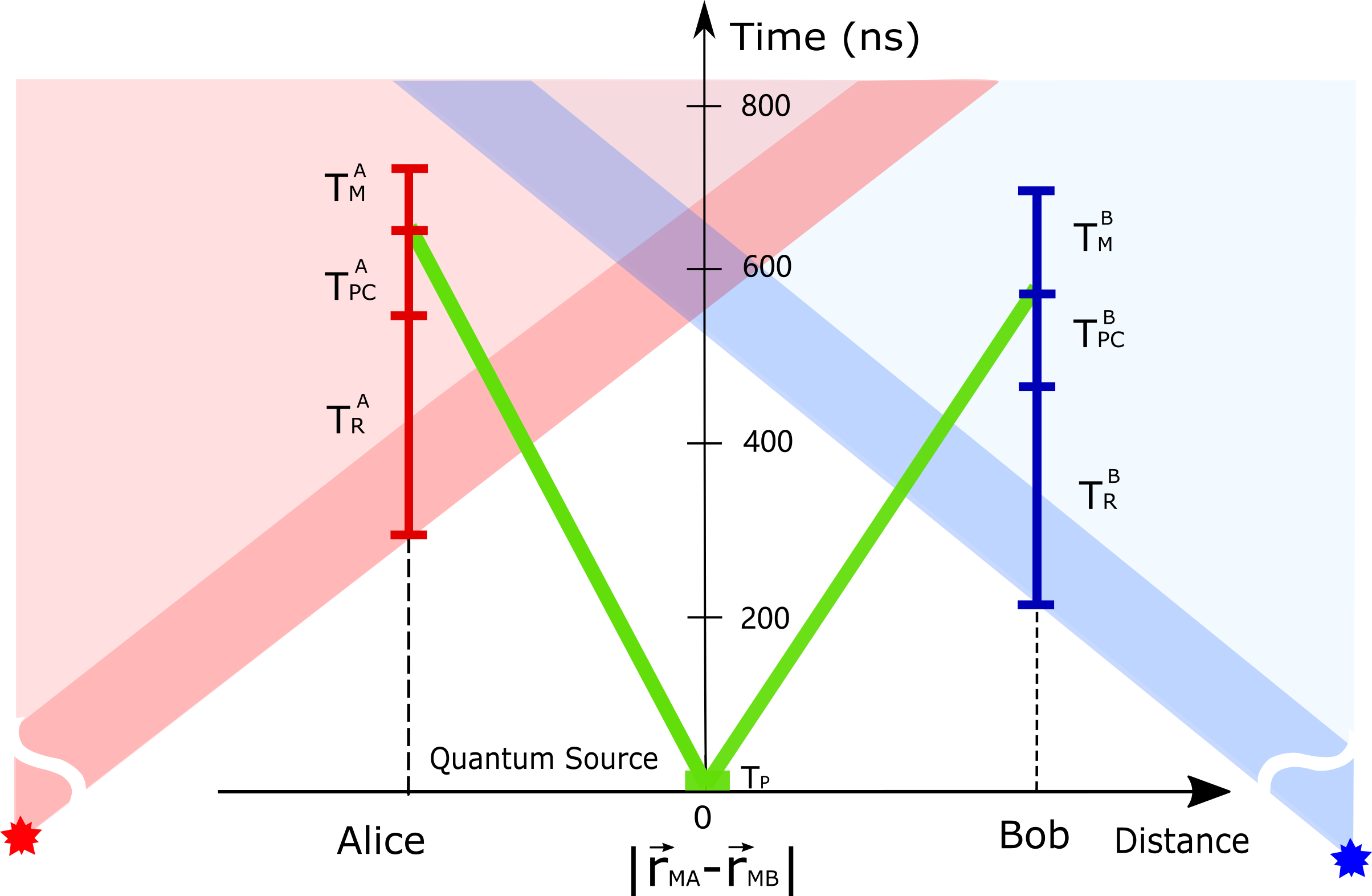}
\caption{Spacetime diagram of the Bell test experiment. The green dot stands for the event of creating an entangled photon pair in the source and the two thick green lines stands for delivering the photons to Alice and Bob via optical fiber, both having an uncertainty of $10$ ns which is the temporal duration of the pump laser pulse. For Alice's (Bob's) side, the red (blue) line segments labeled by $T_R^A$ ($T_R^B$), $T_{PC}^A$ ($T_{PC}^B$), and $T_M^A$ ($T_M^B$)stand for the time elapse starting from cosmic photons arriving the telescope to the Pockels cell receiving a random bit, then to an entangled single photon leaving the Pockels cell, then to the photon detection circuit outputting a signal. The red (blue) strip stands for the time window to accept the cosmic photons for random bit generation satisfying the space-like separation condition through the entire duration of the experiment.}
\label{fig1}
\end{figure*}
where $T_{M}^{A}$ = 55.4 ns ($T_{M}^{B}$ = 99.8 ns) stands for the time elapse starting from an entangled single photon leaving the Pockcels cell to the detection circuit outputting a signal, $T_{P}$ = 10.0 ns is the temporal duration of the pump pulse, $nL_{A}$ = 194 m ($nL_{B}$ = 175 m) is the effective optical path (132 m fiber, 119 m fiber) linking the quantum source and Alice's (Bob's) measurement station, and $c$ is the speed of light in vacuum.
$t_{SA(B)}$ is the time difference between the event of a cosmic source emitting a photon and the event of generating a state in the source, which is given by

\begin{eqnarray}\label{eqb}
  \begin{split}
  t_{SA} = -\frac{|\vec{r}_{SA} - \vec{r}_{MA}|}{c} - T_{R}^A  - T_{PC}^A + \frac{n\cdot L_{A}}{c}, \\
  t_{SB} = -\frac{|\vec{r}_{SB} - \vec{r}_{MA}|}{c} - T_{R}^B  - T_{PC}^B + \frac{n\cdot L_{B}}{c},
  \end{split}
\end{eqnarray}
where $T_{R}^{A}$ = 254 ns ($T_{R}^{B}$ = 249 ns) is the time elapse starting from cosmic photons arriving Alice's (Bob's) telescope to the Pockcels cell receiving a random bit, which includes 26 ns accounting for the refraction of cosmic photons passing through the atmosphere~\cite{stone2011engineering}, $T_{W}=$ 133.2 ns for the allowed time window to receive cosmic photons for generating random bits due to spacelike separation requirement in each experimental run, 43 ns from the telescope receiving a cosmic photon to SPAD outputting a signal, 35 ns for FPGA processing an SPAD signal to generate a random bit, and 17 ns (12 ns) delay for electric cable on Alice's (Bob's) side. $T_{PC}^{A}$ = 111.6 ns ($T_{PC}^{B}$ = 99.2 ns) is the time elapse starting from the Pockcels cell receiving a random bit to an entangled single photon leaving the Pockcels cell.

Defining $\delta \varphi$ as the angular separation between $(\vec{r}_{SA} - \vec{r}_{MA})$ and $(\vec{r}_{SB}-\vec{r}_{MB})$ with
\begin{equation}
  \cos {(\delta \varphi_{A(B)})} = \frac{(\vec{r}_{SA(SB)} - \vec{r}_{MA}) \cdot (\vec{r}_{SA(SB)}-\vec{r}_{MB})} {|\vec{r}_{SA(SB)} - \vec{r}_{MA}| \cdot |\vec{r}_{SA(SB)} - \vec{r}_{MB}|},
\end{equation}
and $\theta_{A(B)}$ as the angular separation between $(\vec{r}_{MA}-\vec{r}_{MB})$ and $(\vec{r}_{SA(SB)} - \vec{r}_{MA(SB)})$ with
\begin{equation}\label{eq}
  \cos {\theta_{A(B)}} = \frac{(\vec{r}_{MA}-\vec{r}_{MB}) \cdot (\vec{r}_{SA(SB)} - \vec{r}_{MA(MB)})} { |\vec{r}_{MA} - \vec{r}_{MB}| \cdot |\vec{r}_{SA(SB)} - \vec{r}_{MA(SB)}|},
\end{equation}
we have
\begin{eqnarray}\label{eq4}
  \begin{split}
  \frac{|\vec{r}_{SA} - \vec{r}_{MB}| - |\vec{r}_{SA} - \vec{r}_{MA}|}{c} = \frac {|\vec{r}_{MA}-\vec{r}_{MB}|}{\sin{\delta \varphi_{A}}}\sin{\theta_{A}} - \frac {|\vec{r}_{MA}-\vec{r}_{MB}|}{\sin{\delta \varphi_{A}}}\sin{(\theta_A - \delta \varphi_{A})} > |\vec{r}_{MA}-\vec{r}_{MB}|\cdot\cos{\theta_A},\\
  \frac{|\vec{r}_{SB} - \vec{r}_{MA}| - |\vec{r}_{SB} - \vec{r}_{MB}|}{c} = \frac {|\vec{r}_{MA}-\vec{r}_{MB}|}{\sin{\delta \varphi_{B}}}\sin{\theta_{B}} - \frac {|\vec{r}_{MA}-\vec{r}_{MB}|}{\sin{\delta \varphi_{B}}}\sin{(\theta_B - \delta \varphi_{B})} > |\vec{r}_{MA}-\vec{r}_{MB}|\cdot\cos{\theta_B},
  \end{split}
\end{eqnarray}
where
$\delta \varphi_{A(B)} \simeq \frac{|\vec{r}_{MA} - \vec{r}_{MB}|}{|\vec{r}_{SA(SB)} - \vec{r}_{MA(MB)}|} \ll 1''$ in our experiment. With Eq.~\ref{eqb} and Eq.~\ref{eq4}, we define parameters $\Gamma^{A}$ and $\Gamma^{B}$ as the difference of the LHS and the RHS of Eq.~\ref{eqa},

\begin{eqnarray}
  \begin{split}
     \Gamma^{A} = \frac{|\vec{r}_{MA} - \vec{r}_{MB}|}{c}\cdot\cos{\theta_A} - T_{P} - T_{R}^A  - T_{PC}^A - T_{M}^B + \frac{n(L_{A} - L_{B})}{c}, \\
     \Gamma^{B} = \frac{|\vec{r}_{MA} - \vec{r}_{MB}|}{c}\cdot\cos{\theta_B} - T_{P} - T_{R}^B  - T_{PC}^B - T_{M}^A + \frac{n(L_{B} - L_{A})}{c}.
  \end{split}
  \label{eq5}
\end{eqnarray}
$\Gamma^{A} > 0$ and $\Gamma^{B} > 0$ indicate satisfying space-like separation in the Bell test experiment. The parameters for Eq.~\ref{eq5} are listed in Table~\ref{tab1}.
\begin{table}
\centering
\caption{Data for latency and distance measurement.}
\begin{tabular}{|c|c|c|c|c|c|c|}
  \hline
  	  k  &$T_P^{k} \pm \sigma T_P^{k} $ (ns) &$T_{R}^{k} \pm \sigma T_{R}^{k}$ (ns)	&$T_{PC}^{k} \pm \sigma T_{PC}^{k} $ (ns)	&$T_{M}^{k}  \pm \sigma T_{M}^{k} $ (ns) &$nL_{k}/c  \pm \sigma_{k} $ (ns)	& $|\vec{r}_{MA}-\vec{r}_{MB}|/c \pm \sigma$ (ns) \\
  \hline
  A (Alice)	&10.0 $\pm$ 0.1& 254 $\pm$ 3 &111.6 $\pm$ 0.25& 55.4 $\pm$ 0.1 &646.7 $\pm$ 0.1 & 610 $\pm$ 3 \\
  B (Bob)&10.0 $\pm$ 0.1& 249 $\pm$ 3 &99.2$\pm$ 0.25&99.8 $\pm$ 0.1 &583.3 $\pm$ 0.1 &\\
  \hline
  \end{tabular}
  \label{tab1}
\end{table}

A schematics of the Bell test experiment with a pair of selected cosmic sources is depicted in Fig.~\ref{fig2}. The angle $\theta_{A(B)}$ changes over time due to Earth rotation, so do $\Gamma^{A}$ and $\Gamma^{B}$. The condition of $\Gamma^{A} > 0$ and $\Gamma^{B} > 0$ can be fulfilled only for a limited period of time for a pair of selected cosmic sources, which can be found based on the star trails of the cosmic source relative to Earth rotation~\cite{Meeus:1991:AA:532892}.
The time period is also restricted due to the unfavorable local condition in Shanghai. To satisfy the spacelike separation requirement and to have enough margin and high SNR in random number generation in the experiment, we set $\theta$ to be between $17^{\circ}$ and $33^{\circ}$. For a smaller $\theta$, the experiment will be subject to serious light pollution. Bigger $\theta$ results in smaller $\Gamma_{A(B)}$. With the choice of $\theta$, by setting the time window $T_{W}=133.2$ ns to collect cosmic photons for random number generation, we calculate $\Gamma_{A}$ and $\Gamma_{B}$ for all experimental runs, with the min ($\Gamma_{A}$, $\Gamma_B$) and the corresponding value for the other star in each run listed in Table~\ref{tab3}.

\begin{table}[htb]
\centering
\caption{The Latitude, Longitude and Elevation of Alice's (Bob's) telescope.}
\begin{tabular}{|c|c|c|c|}
  \hline
          &$Lat.(^{\circ}) $&$Lon.(^{\circ})$&Elev.(m)  \\
  \hline
  Telescope A & N $31.124228^{\circ}$& E $121.547830^{\circ}$&4\\
  Telescope B & N $31.123994^{\circ}$& E $121.545889^{\circ}$&4\\
  \hline
  \end{tabular}
  \label{tab2}
\end{table}

\begin{figure*}[htb]
\centering
\includegraphics [width=8.45 cm,height=2.34 cm]{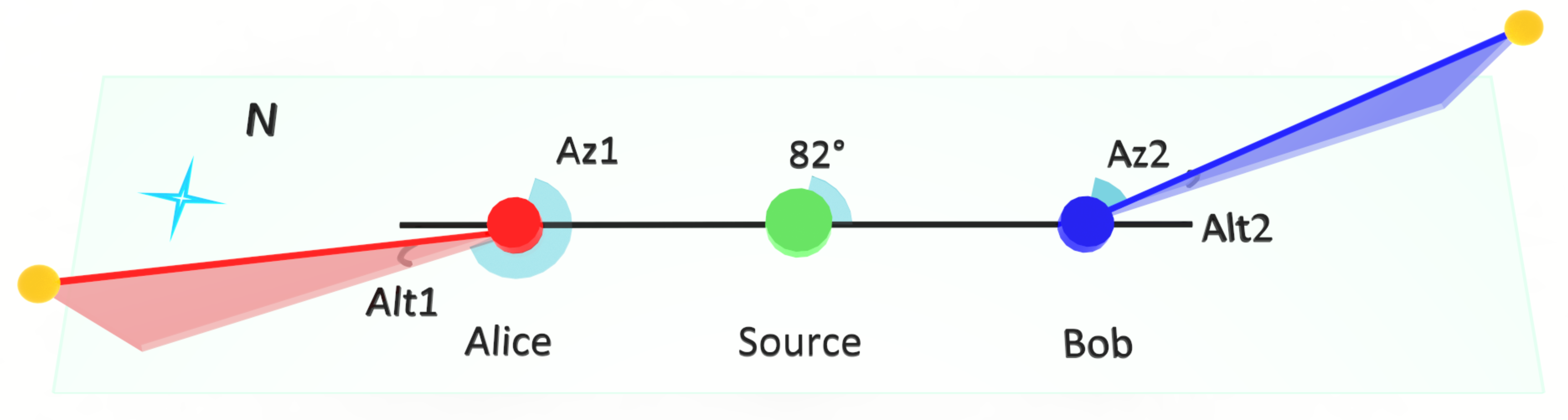}
\caption{A three-dimensional view of the experimental system for an experimental run at a certain time. See Table~\ref{tab3} for the Azimuth (Az) and Altitude (Alt) of each cosmic sources observed at the start time and end time of each experimental run.}
\label{fig2}
\end{figure*}

\begin{table}[htb]
\centering
\caption{More complete version of Table I in the main text. For Alice's(A) and Bob's(B) side, we list Hipparcos ID numbers, celestial coordinates including right ascension and declination, parallax distances ($|\vec{r}_{SA(SB)}|$) with errors ($\sigma_r$) for cosmic sources~\cite{HIP}, the start time and end time of each experimental run on UTC 2018/3/23, azimuth (clockwise from due North, Az) and altitude (Alt) above horizon during the experiment, the angular separation $\alpha$ between Alice's and Bob's cosmic sources, the past light cone intersection event $\tau_{AB}$ with errors ($\delta_{\tau_{AB}}$), and the results of $\Gamma_{A(B)}$ with errors ($\delta_{\Gamma_{A(B)}}=$ 4 ns) at the start time and end time of each experimental run.}

\begin{tabular}{cccccccccccc}
   \hline
   \hline
	run& &HIP ID& $RA^{\circ}$ & $DEC^{\circ}$ & $|\vec{r}_{SA(SB)}| \pm \sigma_r$ (ly) &UTC & $Az^{\circ}$& $Alt^{\circ}$ & $\alpha^{\circ}$ & $\tau_{AB} \pm \delta_{\tau_{AB}} $(yrs)	&$\Gamma_{A(B)}$ (ns) \\
    \hline
    1&	Alice&21421&69.0&16.5&66.63 $\pm$ 0.77&2018/3/23 13h34m&278&20&130.4& 98.99 $\pm$ 0.78&139 \\
		&&&&&&2018/3/23 13h41m&279&18&&&143\\
	&Bob&	69673A&213.9&19.2&36.71 $\pm$	0.22&2018/3/23 13h34m&83&	28&	&&	62 \\
	&&&&&&2018/3/23 13h41m&	84&	29	&&&	56 \\
    \hline
    2&	Alice&	27989&	88.8&7.4&497.86 $\pm$ 63.09&2018/3/23 14h14m&264& 23&131.8&561.81 $\pm$ 62.93&	149 \\
	&&&&&&2018/3/23 14h48m&	269&	16&&								&170 \\
	&Bob&	76267&233.7&26.7&75.03 $\pm$ 0.48&2018/3/23 14h14m&72&23&&&76 \\
	&&&&&&2018/3/23 14h48m&			75&	30	&&&			47 \\
    \hline
    3&	Alice&	37279&114.8&5.22&11.46 $\pm$ 0.05&2018/3/23 15h13m &255&32&	126.5&	148.41 $\pm$ 3.44&	101 \\
	&&&&	&&2018/3/23 15h58m&262&22&&&154 \\
	&Bob& 80816 &247.6&21.5&139.12 $\pm$ 3.44&2018/3/23 15h13m &77&	22&	&&86 \\
	&&&&&&	2018/3/23 15h58m&82&	31&	&&	46\\
    \hline
    4&	Alice&	43813&133.8&5.9&167.15 $\pm$ 1.54&2018/3/23 16h26m &256&33&	126.9&207.91 $\pm$ 1.67&	97 \\
	&&&&&&2018/3/23 17h06m&	262&	24&&&145 \\
	&Bob&	86032	&263.7&12.6&48.58 $\pm$ 0.77&2018/3/23 16h26m &	87&	19&&&	98\\
	&&& &&&2018/3/23 17h06m&		92&	28	&&&				53 \\

 \hline
\end{tabular}
\label{tab3}
\end{table}

\begin{figure}[htb]
  \centering
  \includegraphics[width=9cm]{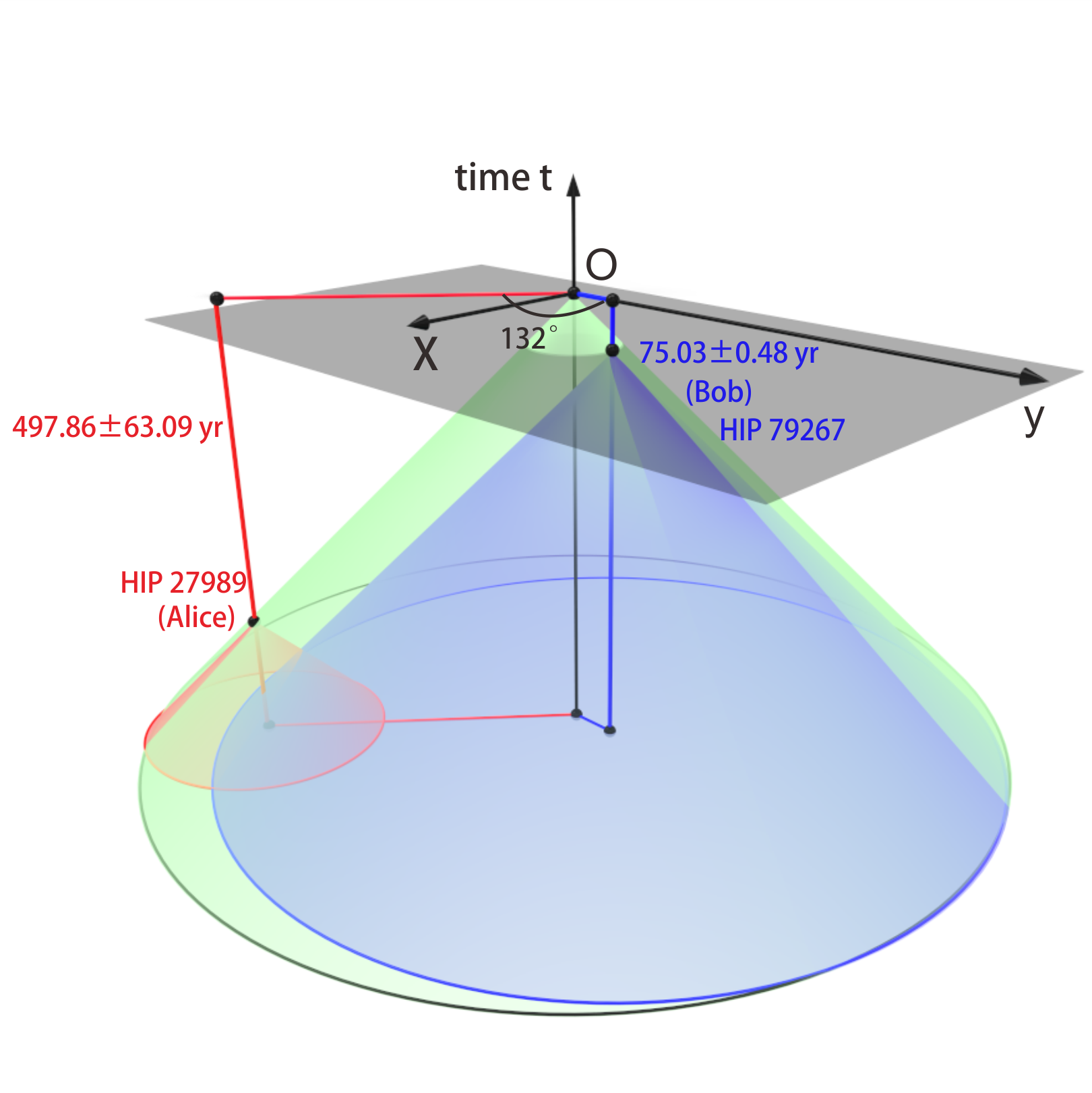}\\
  \caption{Spacetime configuration (shown in $(2+1)$D) in the Bell test experiment using randomness of cosmic photons from remote stars. The past lightcones of events of generating a quantum state in the source (taken as the origin of the coordinates) and emitting a photon in each star cross each other. In this diagram, the selected stars are HIP 27989 for Alice and HIP 76267 for Bob. The common past begins at $t_{cm}=-75.03$~yrs, (labeled by the light grey plane). We test against local hidden variable models taking place after $t_{cm}$.}
  \label{fig3}
\end{figure}

The lightcones of events in a Bell test experiment cross in the past direction, which set the time constraint to local hidden variable models, as shown in Fig.~\ref{fig3}. The angular separation $\alpha$ between Alice's and Bob's selected cosmic sources is listed in Table~\ref{tab3} and the distance between the two cosmic sources is estimated to be $|\vec{r}_{SA} - \vec{r}_{SB}| = \sqrt{|\vec{r}_{SA}|^2 + |\vec{r}_{SB}|^2 - 2|\vec{r}_{SA}| |\vec{r}_{SB}|\cos{\alpha}}$. The lookback time to the past lightcone intersection event $\tau_{AB}$ is obtained to be
\begin{equation}\label{eq6}
  \tau_{AB} = \frac{1}{2}(|\vec{r}_{SA}| + |\vec{r}_{SB}| + |\vec{r}_{SA} - \vec{r}_{SB}|)/c.
\end{equation}

Considering the distance errors $\sigma_{r_{SA}}$ and $\sigma_{r_{SB}}$. The $1\sigma$ lookback time error is given by
\begin{equation}\label{eq7}
  \sigma_{\tau_{AB}} = \frac{1}{2}\frac{\sqrt{(\frac{\sigma_{r_{SA}}}{c})^2(2\tau_{AB}\cdot c - |\vec{r}_{SB}| - |\vec{r}_{SA}| \cos{\alpha})^2 + (\frac{\sigma_{r_{SB}}}{c})^2(2\tau_{AB}\cdot c - |\vec{r}_{SA}| - |\vec{r}_{SA}| \cos{\alpha})^2}}{2\tau_{AB}\cdot c - |\vec{r}_{SA}| - |\vec{r}_{SB}|}
\end{equation}

The result is listed as Table~\ref{tab3}.

\section{Synchronization and random number generation with cosmic photons based on photon arrival time }

 \begin{figure*}[htb]
  \centering
  \includegraphics[width=16cm]{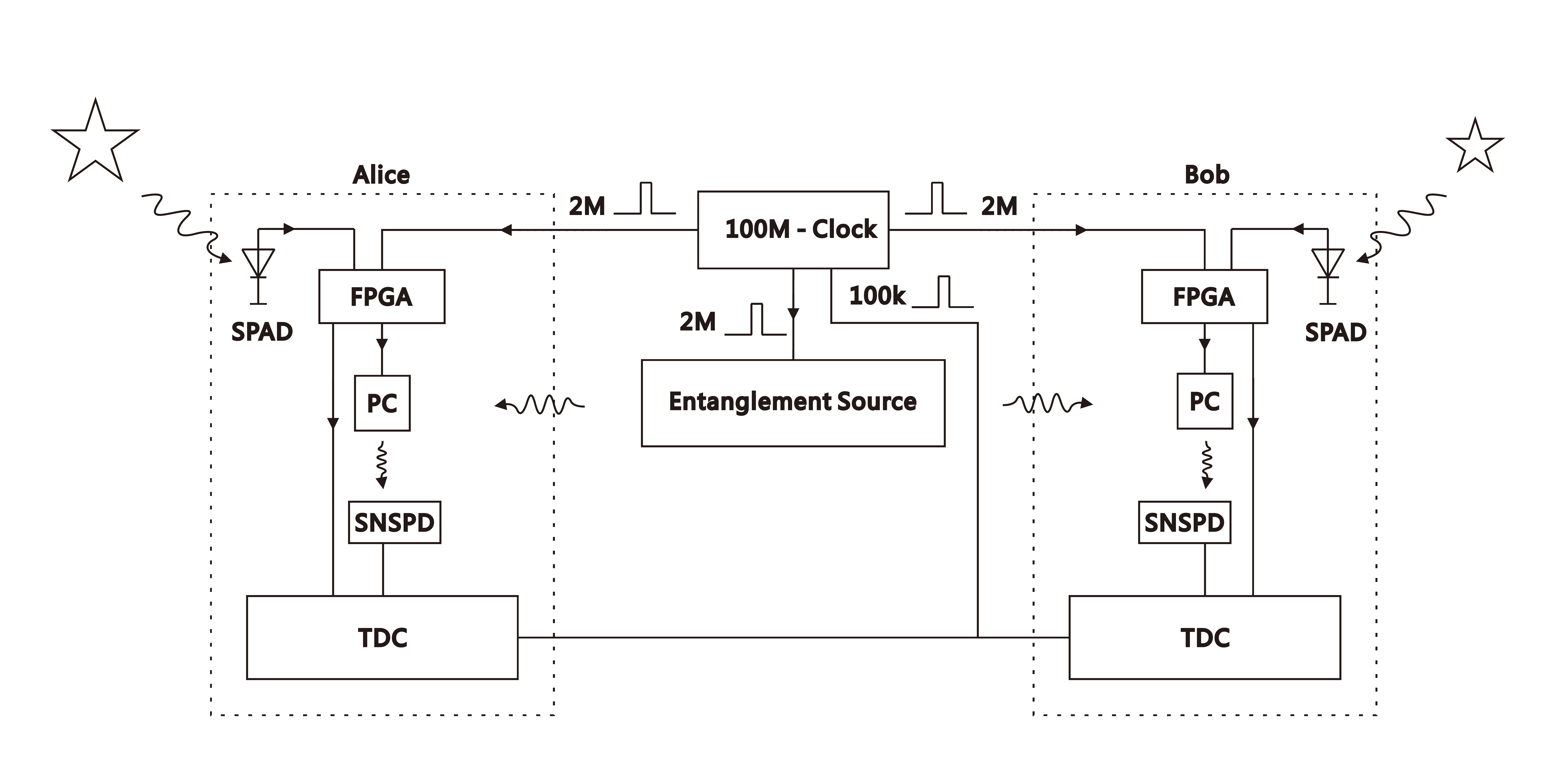}\\
  \caption{Design diagram of the synchronization system.}\label{Fig.sync}
\end{figure*}

A master microwave clock outputs a synchronization signal (SYNC CLK) at a repetition rate of 2 MHz. The SYNC CLK is used to pulse the pump laser to generate entangled photon pairs, and to trigger the field programmable gate array (FPGA). The FPGA outputs a random bit while sending a signal to the time-digital convertor (TDC) for each received photon detection signal. The Pockcels cell prepares the measurement setting upon receiving a random bit from the FPGA to realize the polarization measurement of an entangled single photon from the quantum source, after which, the single photon is detected by a superconducting nanowire single photon detector (SNSPD) with the result passed to TDC. The master clock also sends a synchronization signal at a rate of 100 kHz to synchronize the TDC to time-tag all incoming events.

\begin{figure}[htb]
  \includegraphics[width=16cm]{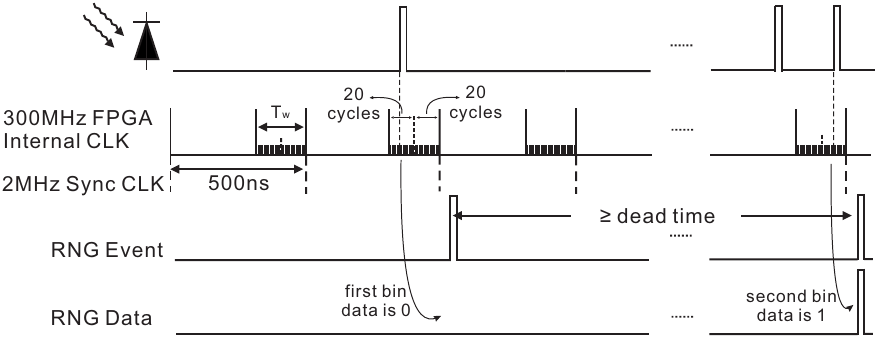}\\
  \caption{Random number generation logic diagram (not drawn to the scale)}
  \label{Fig.RNG}
\end{figure}

A schematics of our random number generation with cosmic photons is depicted in Fig.~\ref{Fig.RNG}. The cosmic photon detection signals from the SPAD are sent to the FPGA which has an internal clock of 300 MHz. The FPGA receives the 2 MHz SYNC CLK from the master clock and the cosmic photon detection signal from the SPAD. The SPAD works in the continuous mode.
The $2$ MHz SYNC CLK and the internal $300$ Mz clock are synchronized: each of the $2$ MHz SYNC CLK is synchronized to one of the $300$ MHz clock signal. Using this as a reference, we count $40$ cycles of the $300$ MHz internal clock before the reference as the time window and divide the window into two time bins. Two bits are used to describe random number generation results. One is RNG event bit, the other is RNG data bit. If the incoming cosmic photon signal falls in the time window, RNG Events outputs ``$1$''. If the incoming cosmic photon detection signal falls in the $1^{st}$ time bin ($1^{st}$ $20$ cycles), RNG Data outputs a bit ``$0$''; if the incoming cosmic photon detection signal falls in the $2^{nd}$ time bin ($2^{nd}$ $20$ cycles), RNG Data outputs a bit ``$1$''. For each successful generation of a random bit, we apply a ``deadtime'' of 5 $\mu$s.

\section{Test of CH-Bell inequality}

The experiment involves two parties, Alice and Bob. Each has an 1-bit output for an 1-bit input in each trial, for a total number of $N$ trials. We consider that the experiment observes the no-signaling theorem. In this case, we consider the CH inequality.  Denote the input and output bits in trial $i$ to be $x_i$ and $y_i$ ($x_i,y_i\in \{0,1\}$) and $a_i$ and $b_i$ ($a_i,b_i\in \{0,1,u\}$), respectively. $u$ denotes the undetected events. The CH inequality can be expressed as a linear combination of the probability distribution $p(ab|xy) $ and given by

\begin{equation}\label{EQ:eberhard}
\begin{aligned}
	J_{CH}=& -p(11|00)-p(11|01)-p(11|10)+p(11|11)\\
&+p^A(1|0)+p^B(1|0),
\end{aligned}
\end{equation}
where $p^A(1|0)$ ($p^B(1|0)$) denotes the probability that the input is 0 and output is 1 in Alice's (Bob's) measurement.
For example more specific, $p^A(1|0)=[p(11|00)+p(10|00)+p(1u|00)+p(11|01)+p(10|01)+p(1u|01)]/2$. Quantum theory allows $J_{CH} < 0$ as opposed to $J_{CH} \geqslant 0$ local hidden variable models.

We configure the experiment in such a way that Alice and Bob's measurement stations are  $ \approx $ 90 meters away from the entanglement source enabling space-like separations between events of state measurements and events of measurement settings, and the measurement settings are arranged to receive inputs from cosmic random number generators (RNGs) considering the freedom-of-choice requirement. The setting parameters are given by the Eberhard's optimization procedure. Meanwhile, we close the fair-sampling loophole with high efficiency photon detection. We repeat the experiment by N trials ($N = \Sigma N_{xy}$), and record the number of correlated events $N_{ab|xy}$ (see Tab.~\ref{tab:Eberhard}). According to Eq.~\eqref{EQ:eberhard}, the $J_{CH}$-value is given by conditional probabilities. The probabilities for settings $x=0,1$, $y=0,1$, and outputs $ab$ are given by
\begin{equation}
	\begin{cases}
	p(11|00) = N_{ab=11|xy=00}/N_{xy=00},\\
	p(11|01) = N_{ab=11|xy=01}/N_{xy=01},\\
	p(11|10) = N_{ab=11|xy=10}/N_{xy=10},\\
	p(11|11) = N_{ab=11|xy=11}/N_{xy=11},\\
	\end{cases}
\end{equation}
where $N_{ab=11|XY}$ are the numbers of events that both Alice and Bob output 1 for base setting choices $X$, $Y$. The other probabilities are obtained similarly.
 For a total number of experimental trials $N = \Sigma N_{xy} = 31448708$, the obtained $J$-value is $-1.405 \times 10^{-4}$, indicating that our system rejects local hidden variable models.

\begin{table}[htb]
\centering
  \caption{Number of correlated events for each measurement base settings: $A_0 B_0$, $A_0 B_1$, $A_1 B_0$ and $A_1 B_1$. $a = 1$ or $0$ indicates that Alice detects a photon or not, the same $b$ for Bob. Mean photon number $\mu=0.07$, violation $J_n=-1.405\times10^{-4}$.}
\begin{tabular}{ccccc}
\hline
Basis settings & $ab=00$ & $ab=10$ & $ab=01$ & $ab=11$\\
\hline
$A_0 B_0$ & 7810453 & 18769 & 18220 & 27895\\
$A_1 B_0$ & 7732172 & 61175 & 11156 & 34477\\
$A_0 B_1$ & 7773073 & 12491 & 62832 & 34234\\
$A_1 B_1$ & 7662717 & 92254 & 93824 & 2966\\
\hline
\end{tabular}
\label{tab:Eberhard}
\end{table}

\section{Data analysis with prediction-based-ratio analysis method}

\subsection{Details of the Prediction-based Ratio (PBR) Analysis}
\label{sect:pbr_test}

Denote Alice's and Bob's random settings at each trial by $X$ and $Y$ with
possible values $x, y\in \{0,1\}$. The setting distributions of Alice
and Bob are $p_{\text{A}}(X)$ and $p_{\text{B}}(Y)$, 
which are assumed to be within total-variance (TV) distances $\epsilon_{\text{A}}$ and
$\epsilon_{\text{B}}$ away from the uniform, respectively. That is,
$|p_{\text{A}}(x)-1/2|\leq \epsilon_{\text{A}}$ and $|p_{\text{B}}(y)-1/2|\leq \epsilon_{\text{B}}$
for all $x,y$. Under the assumption that the setting distributions of Alice and Bob
are independent, their joint-setting distribution is $p(XY)=p_{\text{A}}(X)p_{\text{B}}(Y)$. 
The measurement outcomes of Alice and Bob at each trial are denoted by $A$
and $B$ with possible values $a,b\in \{0,1\}$. To perform a hypothesis test of local realism
using the PBR analysis~\cite{zhang:2011, zhang:2013}, we need to construct a non-negative function
$R$ of the trial result $ABXY$ such that the expectation according to an arbitrary local realistic distribution
$p_{\text{LR}}(AB|XY)$ satisfies
\begin{equation}\label{def:pbr}
\sum_{a,b,x,y} p(xy)p_{\text{LR}}(ab|xy)R(abxy)\leq 1.
\end{equation}
Such a function $R$ is called a PBR. Considering the biases $\epsilon_{\text{A}}$ and
$\epsilon_{\text{B}}$ of Alice's and Bob's setting distributions, the joint-setting distribution
$p(XY)$ is not exactly known. However, as shown in
Ref.~\cite{Knill:2017}, the set of possible distributions $p(XY)$ compatible with the biases
$\epsilon_{\text{A}}$ and $\epsilon_{\text{B}}$ is a convex polytope with only $4$
extreme points. As well known, the set of local realistic distributions is also a convex
polytope (see Refs.~\cite{Fine:1982, Peres:1999} for example), and for the considered Bell-test
configuration the local realistic polytope has $16$ extreme points. Hence, the condition in
Eq.~\eqref{def:pbr} is implied by a finite set of linear constraints on the PBR function
$R$~\cite{Knill:2017}.

Suppose that the experimental distribution at each trial is $p_{\text{exp}}(ABXY)$, then the $p$ value
upper bound according to the PBR analysis decays exponentially when the number of trials $N$ approaches the
infinity, and the decay rate, that is, the confidence-gain rate, is given by
\begin{equation}\label{eq:gain_rate}
g=\sum_{a,b,x,y} p_{\text{exp}}(abxy)\log_{e}(R(abxy)).
\end{equation}
Thus, in order to make the $p$ value upper bound as small as possible, we can
optimize the PBR function $R$ such that the
confidence-gain rate $g$ is as high as possible. The confidence-gain rate
is a concave function of $R$, and the constraints on $R$ in Eq.~\eqref{def:pbr}
are implied by a finite set of linear constraints. Therefore, the optimization
problem for constructing the PBR function $R$ is a convex program and so can
be solved effectively.

However, in practice, the experimental distribution at a trial is unknown. We
estimate it using the observed frequencies $\mathbf{f}\equiv \{n(abxy), a,b,x,y=0,1\}$
where $n(abxy)$ is the number of trials with the result $abxy$.  
We can figure out a no-signaling distribution~\cite{PRBox}
$p^{*}_{\text{NS}}(AB|XY)$ according to which the likelihood of obtaining the observed
frequencies is largest. Since the set of no-signaling distributions is a convex polytope~\cite{Barrett2005}
and the likelihood function is concave, we can find the distribution $p^{*}_{\text{NS}}(AB|XY)$
by convex programming (see Ref.~\cite{zhang:2011} for the details). We just use the
distribution $p^{*}_{\text{NS}}(AB|XY)$ as an estimate of the conditional distribution
of the outcomes $AB$ given the settings $XY$ at a trial. 
If the trial results are independent and identically distributed (i.i.d.) and the number
of trials observed is large, then  $p^{*}_{\text{NS}}(AB|XY)$ will be close to the true
conditional distribution at a trial. In order to estimate the confidence-gain rate, we
can assume that the joint-setting distribution is uniform. Thus, an estimate of the
experimental distribution at a trial is given by $p_{\text{est}}(ABXY)=p^{*}_{\text{NS}}(AB|XY)/4$.
We then construct the PBR function $R$ by maximizing the confidence-gain rate expected
at the estimated distribution $p_{\text{est}}(ABXY)$. We remark that as
long as the non-negative function $R$ satisfies the condition in Eq.~\eqref{def:pbr},
the PBR analysis will provide a valid $p$ value upper bound. We just use the estimated
distribution $p_{\text{est}}(ABXY)$ for helping to construct such functions, but not to
 make a statement about the true distribution of trial results in the experiment.

The above discussion assumes that the trial results are i.i.d. To perform the hypothesis
test without such assumption, before the $i$'th trial we need to construct the PBR function
 $R_i$ for this trial. For this purpose, we need to replace the experimentally
 observed frequencies $\mathbf{f}$ by the frequencies $\mathbf{f}_i$ observed \emph{before}
 the $i$'th trial. The frequencies $\mathbf{f}_i$ can be based on all the trial results before
 the $i$'th trial or using only the most recent trial results in history. (Hence, the
 function $R_i$ is ``prediction-based''.) We also need to replace the biases
 $\epsilon_{\text{A}}$ and $\epsilon_{\text{B}}$ by the biases $\epsilon_{\text{A},i}$
 and $\epsilon_{\text{B},i}$ at the $i$'th trial. Then, following the same procedure as above
 we construct the PBRs  $R_i$, $i=1,2,...,N$. After $N$ trials, the $p$ value upper bound
 for rejecting the null hypothesis of local realism is given by~\cite{zhang:2011}
\begin{equation}
p_{N}=\min\left(\left(\prod_{i=1}^{N}R_i(a_ib_ix_iy_i)\right)^{-1},1\right),
\end{equation}
where $a_ib_i$ and $x_iy_i$ are the experimentally observed outcomes and
setting choices at the $i$'th trial.

The whole experimental running time was divided, according to which star the stellar photons come from,
into four time periods $t_k$, $k=1,2,3,4$. At the time period $t_k$, Alice's and
Bob's measurement stations receive stellar photons from stars $S_{\text{A},k}$ and $S_{\text{B},k}$,
respectively. The random numbers generated by stellar photons are biased, and their
distribution bias changes from a star to another. We estimate the distribution bias
$\epsilon_{\text{A},k}$ or $\epsilon_{\text{B},k}$ of the random numbers coming from
the star $S_{\text{A},k}$ or $S_{\text{B},k}$ (see Table~\ref{tab:bias_estimators} of
Sect.~\ref{sect:bias_estimation} for the details).
We further process the data at each time period $t_k$ block by block. The number of trials in each
data block is around $1.5\times 10^{6}$, and the numbers of data blocks in each time period are
$2$, $5$, $7$, and $7$, respectively.  There are totally $N=31,448,708$ trials in the whole experiment.
We use the same PBR function for all the trials in a data block. This is allowed by the PBR analysis method,
as long as the PBR function for each data block is constructed before processing the data
in this block. For the first data block in each time period $t_k$, we use the trivial PBR function,
that is, $R(abxy)=1$ for all trial results $abxy$. For a latter data block in
the same time period, considering the possible drift of experimental parameters over time, we
construct the PBR function for this block using the frequencies observed in the previous
data block. With considering the distribution biases $\epsilon_{\text{A},k}$ and $\epsilon_{\text{B},k}$,
the PBR analysis returns a $p$ value upper bound $p_{N}=7.873\times 10^{-4}$. If we make the stronger
but unjustified assumption that the joint-setting distribution is perfectly uniform,
the PBR analysis returns a smaller $p$ value upper bound $p'_{N}=3.106\times 10^{-10}$.

\subsection{Test of No Signaling}
\label{sect:ns_test}
The PBR analysis can also be used for the hypothesis test of no signaling. However, if the no-signaling violation by experimental data is not strong, it is possible that the computed $p$ value upper bound with PBRs is not tight.
Either with or without considering the distribution biases $\epsilon_{\text{A},k}$ and $\epsilon_{\text{B},k}$,
the PBR analysis returns a trivial $p$ value upper bound for the hypothesis test of no signaling, suggesting no
obvious evidence of anomalous signaling in the experiment. We also checked whether our experimental data are in agreement with the no-signaling principle by a traditional hypothesis test where the i.i.d. assumption is required.
In the experiment, there are four no-signaling conditions:
the distribution of Alice's outcomes under the setting $x=0$ or $1$ is independent of Bob's setting choices, and the distribution of Bob's outcomes under the setting $y=0$ or $1$ is independent of Alice's setting choices. We performed
a hypothesis test of each no-signaling condition with the two-proportion $Z$-test using the observed frequencies
in the whole experiment (as shown in Table~\ref{tab:Eberhard}). We found the $p$ values of $0.95347$, $0.17608$, $0.37180$, and $0.81156$, which also suggest no obvious evidence of violating the no-signaling principle by our experimental data.

Considering that the observed frequency distribution changes with the time period $t_k$
in which Alice's and Bob's random numbers come from stars $S_{\text{A},k}$ and $S_{\text{B},k}$,
we also performed the hypothesis test of no signaling using the observed frequencies
in each time period $t_k$. At each period $t_k$, we check the four no-signaling
conditions stated in the above paragraph. The $p$ values by the two-proportion $Z$-test assuming
 i.i.d. trial results in each time period are shown in Table~\ref{tab:star_NStest}.
These results suggest no anomalous signaling in each time period of the experiment.

\begin{table}
\caption{$P$ values for the hypothesis test of no signaling in each time period $t_k$,
$k=1,2,3,4$, in which Alice's and Bob's random numbers come from stars $S_{\text{A},k}$
and $S_{\text{B},k}$ respectively. Each column is for one no-signaling condition.}
  \label{tab:star_NStest}
  \begin{equation}
    \begin{array}{|l||l|l|l|l|}
      \hline
      p\text{ value} &x=0&x=1&y=0&y=1\\
      \hline \hline
      k=1 & 0.79623 &   0.33553 &   0.31868 &   0.072244
      \\
      k=2 & 0.86583 &   0.45323 &   0.91513 &   0.88144
      \\
      k=3 & 0.83571 &   0.48486 &   0.54619 &   0.15138
      \\
      k=4 & 0.73539 &   0.63059 &   0.60340 &   0.82749
      \\
      \hline
    \end{array}\notag
  \end{equation}
\end{table}

\subsection{Estimation of Setting Bias}
\label{sect:bias_estimation}
In the experiment, not all random numbers generated at Alice or Bob are used for the Bell test.
In each time period $t_k$, $k=1,2,3,4$, we perform the Bell test only when Alice and Bob
coincidentally detect stellar or background photons.
We can use other detection events at Alice and Bob to estimate the bias of their
own random numbers. In each time period $t_k$, from the raw detection events except their
coincidence detections, Alice and Bob can find the ratios, $r_{\text{A},k}$ and $r_{\text{B},k}$,
of the frequencies of their random numbers $0$ and $1$. They also monitors the signal-to-noise
ratios, $\text{SNR}_{\text{A}, k}$ and $\text{SNR}_{\text{B}, k}$, of the detector clicks
due to stellar photons and those due to background photons in the sky. These results are shown
in Table~\ref{tab:bias_noise_statistics}.

\begin{table}
\caption{Frequency ratios of random numbers $0$ and $1$ and SNRs in each time period $t_k$,
$k=1,2,3,4$.}
  \label{tab:bias_noise_statistics}
  \begin{equation}
    \begin{array}{|l||l|l|l|l|}
      \hline
       k & 1 & 2 & 3 &4\\
      \hline \hline
      r_{\text{A},k} & 1.0078 & 1.0053 & 1.0059 & 1.0009
      \\
      r_{\text{B},k} & 1.0012 & 0.9989 & 0.9985 & 0.9987
      \\
      \text{SNR}_{\text{A}, k} & 491.6 &  584.6 & 147.4 & 125.2
      \\
      \text{SNR}_{\text{B}, k} & 94.7  & 111.6 &  78.5  & 132.6
      \\
      \hline
    \end{array}\notag
  \end{equation}
\end{table}

Let us estimate the bias of Alice's random numbers first. We make the following
two assumptions: 1) The distribution of random numbers $p_{\text{A},k}^{(s)}(X)$,
due to detections of stellar photons in a time period $t_k$, is characterized
by the frequency ratio $r_{\text{A},k}$ 
in the following sense:
\begin{equation}
\max_{x\in \{0, 1\}} p_{\text{A},k}^{(s)}(x)\leq \max\left(\frac{r_{\text{A},k}}{1+r_{\text{A},k}},
\frac{1}{1+r_{\text{A},k}}\right). \label{eq:Alice_star_bias}
\end{equation}
2) The distribution of random numbers $p_{\text{A},k}(X)$, due to detections of either
stellar or background photons in a time period $t_k$, is given by a probabilistic
mixture of the distribution $p_{\text{A},k}^{(s)}(X)$ due to stellar photons and an arbitrary
\emph{uncharacterized} distribution due to background photons, where the mixture weight of
the distribution $p_{\text{A},k}^{(s)}(X)$ is given by
$\frac{\text{SNR}_{\text{A}, k}}{1+\text{SNR}_{\text{A}, k}}$.
Under the above two assumptions, we obtain that 
\begin{align}
\max_{x\in \{0, 1\}} p_{\text{A},k}(x) & \leq \frac{\text{SNR}_{\text{A}, k}}{1+\text{SNR}_{\text{A}, k}}
\max_{x\in \{0, 1\}} p_{\text{A},k}^{(s)}(x) + \frac{1}{1+\text{SNR}_{\text{A}, k}} \notag \\
& \leq \frac{\text{SNR}_{\text{A}, k}}{1+\text{SNR}_{\text{A}, k}} \max\left(\frac{r_{\text{A},k}}{1+r_{\text{A},k}}, \frac{1}{1+r_{\text{A},k}} \right)+\frac{1}{1+\text{SNR}_{\text{A}, k}}.
\end{align}
Therefore, the bias $\epsilon_{\text{A},k}$, which is an upper bound
on $|p_{\text{A},k}(x)-1/2|$ for all $x$, is given by
\begin{equation}
\epsilon_{\text{A},k}= \frac{\text{SNR}_{\text{A}, k}}{1+\text{SNR}_{\text{A}, k}} \max\left(\frac{r_{\text{A},k}}{1+r_{\text{A},k}}, \frac{1}{1+r_{\text{A},k}} \right)+\frac{1}{1+\text{SNR}_{\text{A}, k}}-\frac{1}{2}.
\label{eq:Alice_bias}
\end{equation}
In the same way, we estimate the bias of Bob's random numbers in a time period $t_k$ as
\begin{equation}
\epsilon_{\text{B},k}= \frac{\text{SNR}_{\text{B}, k}}{1+\text{SNR}_{\text{B}, k}} \max \left(\frac{r_{\text{B},k}}{1+r_{\text{B},k}}, \frac{1}{1+r_{\text{B},k}} \right)+\frac{1}{1+\text{SNR}_{\text{B}, k}}-\frac{1}{2}.
\label{eq:Bob_bias}
\end{equation}
Accordingly, the biases estimated using the measurement results in Table~\ref{tab:bias_noise_statistics}
are shown in Table~\ref{tab:bias_estimators}. 

\begin{table}
\caption{Estimated biases of random numbers $0$ and $1$ in each time period $t_k$, $k=1,2,3,4$.}
  \label{tab:bias_estimators}
  \begin{equation}
    \begin{array}{|l||l|l|l|l|}
      \hline
       k & 1 & 2 & 3 &4\\
      \hline \hline
      \epsilon_{\text{A},k} & 0.00295 & 0.00217 & 0.00483 & 0.00419
      \\
      \epsilon_{\text{B},k} & 0.00552 & 0.00471 & 0.00666 & 0.00407
      \\
      \hline
    \end{array}\notag
  \end{equation}
\end{table}

We also verified that in each time period $t_k$ the observed frequencies of random numbers
$0$ and $1$ in the Bell-test trials where Alice and Bob coincidentally detect stellar or
background photons are consistent with the estimated biases $\epsilon_{\text{A},k}$
and $\epsilon_{\text{B},k}$.  Let the number of Bell-test trials with Alice's input $x$,
$x=0,1$, in a time period $t_k$ be $n_{\text{A},k}(x)$. 
Then, the consistency of Alice's random inputs with the estimated biases 
$\epsilon_{\text{A},k}$ is suggested if the probability
\begin{equation}
c_{\text{A},k} \equiv \text{Prob}\left(\frac{\max_{x} n_{\text{A},k}(x)}{n_{\text{A},k}(0)+n_{\text{A},k}(1)}
\leq \frac{1}{2}+\epsilon_{\text{A},k}\right),
\label{eq:Alice_consitency}
\end{equation}
takes a large value.
In parallel, let the number of Bell-test trials with Bob's input $y$, $y=0,1$, in a time period $t_k$
be $n_{\text{B},k}(y)$. Then, the consistency at Bob's side is suggested by a large value
of the probability
\begin{equation}
c_{\text{B},k} \equiv \text{Prob}\left(\frac{\max_{y} n_{\text{B},k}(y)}{n_{\text{B},k}(0)+n_{\text{B},k}(1)}
\leq \frac{1}{2}+\epsilon_{\text{B},k}\right).
\label{eq:Bob_consistency}
\end{equation}
By Hoeffding's bound~\cite{Hoeffding} with the observed frequencies shown in Table.~\ref{tab:consistency}, we can lower
bound the two probabilities in Eqs.~\eqref{eq:Alice_consitency}
and~\eqref{eq:Bob_consistency}. The corresponding lower bounds are also shown
in Table.~\ref{tab:consistency}.

\begin{table}
\caption{Consistency check of the estimated biases $\epsilon_{\text{A},k}$ and $\epsilon_{\text{B},k}$
 in each time period $t_k$, $k=1,2,3,4$.}
  \label{tab:consistency}
  \begin{equation}
    \begin{array}{|l||l|l|l|l|}
      \hline
       k & 1 & 2 & 3 &4\\
      \hline \hline
      n_{\text{A},k}(0) & 1,486,191 & 4,072,171 & 5,382,580  & 4,817,025
      \\
      n_{\text{A},k}(1) &  1,478,591 & 4,048,210 & 5,351,255 & 4,812,685
      \\
      c_{\text{A},k} & \geq (1-6.35\times 10^{-8}) & \geq 0.99963 & \geq (1-1.14\times 10^{-106})  &
      \geq (1-7.04\times 10^{-132})
      \\
      \hline
       n_{\text{B},k}(0) & 1,481,833 & 4,056,701 & 5,364,939 & 4,810,844
      \\
      n_{\text{B},k}(1) &  1,482,949 & 4,063,680 & 5,368,896 & 4,818,866
      \\
      c_{\text{B},k} & \geq (1-10^{-73}) & \geq (1-10^{-129}) & \geq (1-10^{-390})  &
      \geq (1-10^{-111})
      \\
      \hline
    \end{array}\notag
  \end{equation}
\end{table}

\end{document}